# Space nutrition: the key role of nutrition in human space flight

Catalano Enrico[1]

[1]Consorzio Interuniversitario Nazionale per la Scienza e Tecnologia dei Materiali (INSTM), Via Giuseppe Giusti, 9, 50121 Firenze (FI), Italy. e-mail: enrico.catalano@med.uniupo.it

**Corresponding author**

Catalano Enrico

Consorzio Interuniversitario Nazionale per la Scienza e Tecnologia dei Materiali (INSTM), Via Giuseppe Giusti, 9, 50121 Firenze (FI), Italy. e-mail: enrico.catalano@med.uniupo.it

**Abstract**

From the basic impact of nutrient intake on health maintenance to the psychosocial benefits of mealtime, great advancements in nutritional sciences for support of human space travel have occurred over the past 50 years. Nutrition in space has many areas of impact, including provision of required nutrients and maintenance of endocrine, immune, and musculoskeletal systems. It is affected by environmental conditions such as radiation, temperature, and atmospheric pressures, and these are reviewed. Nutrition with respect to space flight is closely interconnected with other life sciences research disciplines including the study of hematology, immunology, as well as neurosensory, cardiovascular, gastrointestinal, circadian rhythms, and musculoskeletal physiology. Psychosocial aspects of nutrition are also important for more productive missions and crew morale. Research conducted to determine the impact of spaceflight on human physiology and subsequent nutritional requirements will also have direct and indirect applications in Earth-based nutrition research. Cumulative nutritional research over the past five decades has resulted in the current nutritional requirements for astronauts. Realization of the full role of nutrition during spaceflight is critical for the success of extended-duration missions. Long-duration missions will require quantitation of nutrient requirements for maintenance of health and protection against the effects of microgravity.

## 1. Introduction

Nutrition has played a critical role throughout the history of exploration, and space exploration is no exception. The purpose of this article is to review our current knowledge of space flight nutrition and food science. Humans have adapted well to space flight, and over the past 50 years, we have substantially increased our understanding of the various physiologic changes that occur during and after space flight [1]. Nutrition in space has many aspects of impact, including provision of required nutrients and maintenance of endocrine, immune, and musculoskeletal systems. However, the underlying mechanisms for many of these alterations remain unclear. This article describes prior and ongoing nutritional research undertaken with the goal of assuring human health and survival during space flight. Nutrition and food science research overlap with or are integral to many other aspects of space medicine and physiology including psychological health, sleep and circadian rhythmicity, taste and odor sensitivities, radiation exposure, body fluid shifts, and wound healing and to changes in the musculoskeletal, neurosensory, gastrointestinal, hematologic, and immunologic systems. Nutrient intake play a fundamental role from health maintenance to the psychosocial benefits of mealtime, the role of nutrition in space is evident. Recent advances in genomics and proteomics are just beginning to be applied in space biomedical research, and it is likely that findings from such studies will be applicable to applied human nutritional science.

Long-duration missions will require the right amount of nutrient requirements for maintenance of health and protection against the effects of microgravity. Psychosocial aspects of nutrition will also be important for more productive missions and crew morale. Realization of the full role of nutrition during spaceflight is critical for the success of extended-duration missions. Research conducted to determine the impact of spaceflight on human physiology and subsequent nutritional requirements will also have direct and indirect applications in Earth-based nutrition research.

Understanding the nutritional requirements of space travelers and the role of nutrition in human adaptation to microgravity are as critical to crew safety and mission success as any of the mechanical systems of the spacecraft itself. There are many facets to nutrition and health on Earth. Space flight introduces further complications, and many gaps remain in our knowledge of the relationships between nutrition and health in space that need to be filled before we can safely embark on "exploration" missions, that is, missions beyond low Earth orbit. At the surface of these unknowns is the need to understand and define basic nutrient requirements during extended stays in microgravity. Beyond this lies the need to characterize the role of nutrition in the adaptation of physiological systems to microgravity, and/or the impact of these changes on nutrition. Additionally, environmental impacts (including radiation, and spacecraft and spacesuit atmospheres) can alter nutritional status and nutritional requirements of space flight. For surface missions (on, for example, the moon and Mars), partial gravity may complicate the situation further. Finally, many potential targets for nutritional countermeasures exist, where modified dietary intake may help to counteract or mitigate some of the negative effects of space flight on the human body.

The US space life sciences research community has developed a set of critical questions and a road map to clearly emphasize research efforts that ultimately will reduce to humans the risk associated with space travel and habitation [2]. Relevant research has been conducted in space and on the ground using animal models and human ground-based analogs [3]. Throughout the five-decade history of human space flight, nutrition and food research have been an integral component of various missions.

## 2. Astronaut nutrition

Spacecraft, the space environment, and weightlessness itself all impact human physiology. Clean air, drinkable water, and effective waste collection systems are required for maintaining a habitable environment. Adequate energy intake is perhaps the single most important aspect of astronaut nutrition [4]. This is not only because energy in and of itself is more important than other nutritional factors, but also because if enough food is consumed to meet energy needs, then generally other nutrients will also be consumed in reasonable amounts. Weightlessness impacts almost every system in the body, including those of the bones, muscles, heart and blood vessels, and nerves. There are many facets to maintaining eucaloric intake during space flight, including energy requirements; physiological changes in taste and satiety; scheduling issues of allotting time for meal preparation, consumption, and cleanup; food quality; and even food availability [4]. Little research has been done on differences in fuel components (protein, carbohydrate, fat) during space flight, or on cofactors (eg, vitamins) of energy utilization. We review these here, highlighting what has been done and potential areas of future research.

### 2.1 Space food

Space food is a variety of food products, specially created and processed for consumption by astronauts in outer space. The food has specific requirements of providing balanced nutrition for individuals working in space, while being easy and safe to store, prepare and consume in the machinery-filled low gravity environments of manned spacecraft [1]. In recent years, space food has been used by various nations engaging on space programs as a way to share and show off their cultural identity and facilitate intercultural communication. Although astronauts consume a wide variety of foods and beverages in space, the initial idea was to supply astronauts with a formula diet that would supply all the needed vitamins and nutrients [2]. With the long-duration missions aboard the International Space Station (ISS), it has become clear that more emphasis needs to be placed on improving human habitability. In fact, in the last years a new project: "The Vegetable Production System" (VEGGIE) was developed to provide a means to supply crews with a continuous source of fresh food and a tool for relaxation and recreation. The Veggie provides lighting and nutrient delivery, but utilizes the cabin environment for temperature control and as a source of carbon dioxide to promote growth [64].

The Vegetable Production System (Veggie) is a deployable plant growth unit capable of producing salad-type crops to provide the crew with a palatable, nutritious, and safe source of fresh food and a tool to support relaxation and recreation. During 2015 red romaine lettuce was cultivated and consumed on board on ISS by the crew members [65]. Veggie will remain on the station permanently and could become a research platform for other top-growing plant experiments.

**3. Bone and nutrition in space**

Bone is a living tissue, and is constantly being remodeled. This remodeling is achieved through breakdown of existing bone tissue (a process called resorption) and formation of new bone tissue. Chemicals in the blood and urine can be measured to determine the relative amounts of bone resorption and formation. During spaceflight, bone resorption increases significantly, and formation either remains unchanged or decreases slightly [5]. The net effect of this imbalance is a loss of bone mass. Bone loss, especially in the legs, is increased during spaceflight. This is most important on flights longer than thirty days, because the amount of bone lost increases as the length of time in space increases. Weightlessness also increases excretion of calcium in the urine and the risk of forming kidney stones. Both of these conditions are related to bone loss [5].

Calcium and bone metabolism remain key concerns for space travelers, and ground-based models of space flight have provided a vast literature to complement the smaller set of reports from flight studies. Increased bone resorption and largely unchanged bone formation result in the loss of calcium and bone mineral during space flight, which alters the endocrine regulation of calcium metabolism [6]. Physical, pharmacologic, and nutritional means have been used to counteract these changes. In 2012, heavy resistance exercise plus good nutritional and vitamin D status were demonstrated to reduce loss of bone mineral density on long-duration International Space Station missions [6].

Uncertainty continues to exist, however, as to whether the bone is as strong after flight as it was before flight and whether nutritional and exercise prescriptions can be optimized during space flight. Findings from these studies not only will help future space explorers but also will broaden our understanding of the regulation of bone and calcium homeostasis on Earth.

Whereas nutrition is critical for virtually all systems, the interaction of nutrition with bone is perhaps more extensive and complex than most of its interactions with body systems. Many nutrients are important for healthy bone, particularly calcium and vitamin D. When a food containing calcium is eaten, the calcium is absorbed by the intestines and goes into the bloodstream. Absorption of calcium from the intestines decreases during spaceflight. Even when astronauts take extra calcium as a supplement, they still lose bone [7].

Bone is the body's reservoir of calcium, which provides structure and strength to bone, but also provides a ready resource to maintain blood calcium levels during periods with insufficient dietary provision of calcium. Several nutrients are required for the synthesis of bone, including protein and vitamins D, K, and C. On Earth, the body can produce vitamin D after the skin is exposed to the sun's ultraviolet light. In space, astronauts could receive too much ultraviolet light, so spacecraft are shielded to prevent this exposure. Because of this, all of the astronauts' vitamin D has to be provided by their diet. However, it is very common for vitamin D levels to decrease during spaceflight [7].

Multiple risks are associated with bone loss during space flight. Almost immediately upon entering weightlessness, bone resorption increases, and calcium (and other minerals) are released into the blood and urine. This increases kidney stone risk on short missions, and on longer missions, chronic bone and calcium loss can increase risks to bone health both in the near term (eg, risk of fractures) and in the long term (eg, risk of osteoporosis-like bone degradation).

Sodium intake is also a concern during spaceflight, because space diets tend to have relatively high amounts of sodium. Increased dietary sodium is associated with increased amounts of calcium in the urine and may relate to the increased risk of kidney stones. The potential effect of these and other nutrients on the maintenance of bone health during spaceflight highlights the importance of optimal dietary intake. The changes in bone during spaceflight are very similar to those seen in certain situations on the ground [7]. There are similarities to osteoporosis, and even paralysis. While osteoporosis has many causes, the end result seems to be similar to spaceflight bone loss. Paralyzed individuals have biochemical changes very similar to those of astronauts. This is because in both cases the bones are not being used for support. In fact, one of the ways spaceflight bone loss is studied is to have people lie in bed for several weeks. Using this approach, scientists attempt to understand the mechanisms of bone loss and to test ways to counteract it. If they can find ways to successfully counteract spaceflight bone loss, doctors may be able to use similar methods to treat people with osteoporosis or paralysis. It is not clear whether bone mass lost in space is fully replaced after returning to Earth. It is also unclear whether the quality (or strength) of the replaced bone is the same as the bone that was there before a spaceflight. Preliminary data seem to show that some crew members do indeed regain their preflight bone mass, but this process takes about two or three times as long as their flight. The ability to understand and counteract weightlessness-induced bone loss remains a critical issue for astronaut health and safety. The International Space Station provides the opportunity for monitoring the effects of space flight on bone physiology to be documented, along with the testing of countermeasures aimed at counteracting bone loss [8].

## 4. Muscle loss in microgravity conditions

Loss of body weight (mass) is a consistent finding throughout the history of spaceflight. Typically, these losses are small (1 percent to 5 percent of body mass), but they can reach 10 percent to 15 percent of preflight body mass. Although a 1 percent body-weight loss can be explained by loss of body water, most of the observed loss of body weight is accounted for by loss of muscle and adipose (fat) tissue. Exposure to microgravity reduces muscle mass, volume, and performance, especially in the legs, on both short and long flights [9-11].

Weightlessness leads to loss of muscle mass and muscle volume, weakening muscle performance, especially in the legs. The loss is believed to be related to a metabolic stress associated with spaceflight. These findings are similar to those found in patients with serious diseases or trauma, such as burn patients.

The primary countermeasure against muscle loss remains adequate energy intake, which will no doubt include protein, but protein supplements are not required. Use of protein and amino acid supplementation has long been studied as a potential means to mitigate muscle loss associated with space flight, but results have been inconclusive at best [12, 13].

It remains unclear whether nutritional measures beyond the consumption of adequate energy and protein would be beneficial in reducing muscle atrophy [14].

Exercise routines have not succeeded in maintaining muscle mass or strength of astronauts during spaceflight. Most of the exercises performed have been aerobic (e.g., treadmill, stationary bicycle). Use of resistance exercise, in which a weight (or another person) provides resistance to exercise against, has been proposed to aid in the maintenance of both muscle and bone during flight. Ground-based studies (not done in space) of resistance exercise show that it may be helpful, not only for muscle but also for bone [8]. Studies being conducted on the International Space Station are testing the effectiveness of this type of exercise for astronauts.

Exercise is the most common first-pass approach to maintaining muscle mass and strength [15]. The exercise regimens tested as countermeasures to date have generally not succeeded in maintaining muscle mass or strength (or bone mass) during space flight [15-17]. Many types of exercise protocols have been proposed to aid in the maintenance of both muscle and bone during flight, but these have yet to be fully tested on orbit [18, 19].

## 5. Endocrine system function related to the nutritional status in the space flight

The endocrine system appears to be sensitive to the conditions of space flight. Several hormones may increase in the circulation as part of the stress response to microgravity conditions, including epinephrine and norepinephrine, adrenocorticotropin, cortisol. These hormones play a role in the elevation of plasma glucose and fatty acids, in increased lipolytic activity in adipose tissue, in reducing lipogenesis and in raising glycogen content in the liver [20]. These hormones have been measured during and after space flight: data from previous space missions show increases in catabolic hormones (cortisol, glucagone) and a prolonged elevation of 3-methylhistidine excretion, suggesting a chronic metabolic stress response that may be influenced not only by mission length but also by energy intake, exercise regimen and even gender [21]. Such changes may indeed be involved in promoting muscle and bone loss, in impairing immune status, in the regulation of body fluids and electrolytes that affects the cardiovascular response to microgravity. Catecholamines as well as renin-aldosterone are also involved in fluid balance, being part of sodium-retaining endocrine systems; consistent observations in various missions (Mir 97, Spacelab mission D-2, Euromir 94) revealed an elevated activity that may lead to sodium storage without an accompanying fluid retention [22]. This may be part of the reason leading to an extravasation of fluid after an increase in vascular permeability. An enormous capacity for sodium in the extra-vascular space and a mechanism that allows the dissociation between water and sodium handling may contribute to fluid balance adaptation in weightlessness [23].

## 6. Gastrointestinal function during space missions

Astronauts experience gastrointestinal changes early in flight, gaseous stomach occurs due to the inability of gases to rise. Furthermore the effects of micro-gravity are presumed to alter the contact of the gastric contents with the gastrointestinal mucus. However, cephalic fluid shifts, in combination with commonly observed dehydration, could possibly affect gastrointestinal motility through reduced splanchnic flow. The effect of chronic inactivity increases transit time and potentially changes gastrointestinal microflora [24]. Gastrointestinal integrity and bacterial balance may be improved by probiotics and prebiotics that should be studied for a possible inclusion in space foods. Probiotics, defined as microbial food supplements that beneficially affect the host improving its intestinal microbial balance, have been studied to change the composition of colonic microbiota by increasing bacterial groups such as Bifidobacteria and Lactobacilli that are perceived as exerting health-promoting properties. However, these changes may be transient, and the implantation of exogenous bacteria becomes limited. On the other hand, since astronauts do have an impaired immune function, as discussed in the next section, differences in the immunomodulatory effects of candidate probiotic bacteria should be taken into consideration and the positive effect should be studied in weightlessness models. The use of prebiotics partly overcomes the limitations of probiotics. Prebiotics are growth substrates and not viable entities, specifically directed toward potentially beneficial bacteria already growing in the colon [25].

## 7. Cardiovascular health and nutrition in space

Cardiovascular issues are a key concern for space travelers, but the role of nutrition in cardiovascular adaptation has not yet been well characterized [26-28]. It is worth noting that multiple studies are being planned or are underway on ISS to shed light on this area in the near future. Omega-3 fatty acids have a clear beneficial impact on cardiovascular health on Earth, but such effects have not been evaluated during space flight [29]. Nonetheless, the initial efforts being made to increase fish and omega-3 fatty acid intake in astronauts for the benefit of other systems (bone, muscle) will likely have positive effects here as well [29].

## 8. Iron metabolism during space flight

Space flight-induced hematologic changes have been observed since the initial days of space exploration. Space flight anemia is a widely recognized phenomenon in astronauts. The implications of hematologic changes for long-term space flight may have consequences on the health of astronauts [30]. Reduction in circulating red blood cells and plasma volume results in a 10% to 15% decrement in circulatory volume. This effect appears to be a normal physiologic adaptation to weightlessness and results from the removal of newly released blood cells from the circulation [30]. Iron availability increases, and (in the few subjects studied) iron stores increase during long-duration space flight. The consequences of these changes are not fully understood.

## 9. Oxidative stress and space flight

Space flight is associated with an increase in oxidative stress and to significantly high radiation, even with considerable shielding on the spacecraft. In the human body, solar radiation or low wavelength electromagnetic radiation (such as gamma rays) from the earth or space environment can split water to generate reactive free radicals. These reactive free radicals can react in the body leading to oxidative damage to lipids, proteins and DNA. Recent data on the oxidant damage have underlined its increase post-flight probably due to a combination of augmented metabolic activity and loss of some host antioxidant defences in-flight [31, 32].

The effect is more pronounced after long-duration space flight. The effects last for several weeks after landing. In humans there is increased lipid peroxidation in erythrocyte membranes, reduction in some blood antioxidants, and increased urinary excretion of markers for oxidative damage to lipids and DNA, respectively: 8-iso-prostaglandin F2α and 8-oxo-7,8 dihydro-2 deoxyguanosine [31, 33].

Decreasing the imbalance between the production of endogenous oxidant defenses and oxidant production by increasing the supply of dietary antioxidants may lessen the severity of the postflight increase in oxidative stress. The antioxidant defence system can be implemented to counteract oxidative stress by supplementing vitamins in dietary intake of astronauts such as tocopherols and tocotrienols (vitamin E), ascorbate (vitamin C), vitamin A and its precursors betacarotene and other carotenoids, trace elements and minerals such as copper, manganese, zinc, selenium and iron [34]. Actual recommendations of vitamin C intake have been raised from 60 to 100 mg per day, during space flights [4, 35]. Dietary and antioxidant defences appear to play a protective role in muscle cells by reducing associated oxidative damage to lipids, nucleic acids, and proteins [34]. However, iron supplementation in microgravity is not recommended because the reduction in red cell mass and the consequent increase in iron stores could augment free radical generation.

Dietary intake and/or supplementation of particular nutrients showed also an important prophylactic role against photo-oxidative damage to cell membranes [36]. Nutritional countermeasures in space with antioxidant supplementation provide a great opportunity for research within space flight models, whereas the main concern is to counteract the potential for long-term oxidative damage in humans under these conditions.

## 10. Ophthalmic changes in space flight

Microgravity is the dominant cause of many physiological changes during spaceflight and is thought to contribute significantly to the observed ophthalmic changes. Ophthalmic health among astronauts has gotten attention in recent years because of a newly identified issue for some crewmembers. In addition to a general increase in cataract risk, some crewmembers have experienced vision-related changes after long-duration space flight [37, 38].

It has recently been recognized that vision changes are actually quite common in astronauts and are associated with a constellation of findings including elevated intracranial pressure, optic disc edema, globe flattening, optic nerve sheath thickening, hyperopic shifts, choroidal folds, cotton wool spots and retinal changes [39]. The etiology of the refractive and structural ophthalmic changes is currently not known and continues to be researched, but biochemical evidence indicates that the folate- and vitamin B12-dependent 1-carbon transfer pathway may be involved. Nutrition is known to be an important factor for ophthalmic health in general.

## 11. Immune changes in space flight

Optimal function of the immune system is impaired in the presence of malnutrition. Without adequate nutrition, the immune system is clearly deprived of the components needed to generate an effective immune response [40, 41]. Nutrients act as antioxidants and as cofactors [42]. Exposure of animals and humans to space flight conditions has resulted in numerous alterations in immunological parameters, for instance, decreases in blast transformation of lymphocytes, cytokine production, and natural killer cell activity. After flight, alterations in leukocyte subset distribution have also been reported for humans and animals [43]. Disruption of nutritional balance and dietary intake of astronauts and cosmonauts during space flight, which is often accompanied by a stress response, might influence their immune response [44, 45]. Findings of energy deficiency on long duration missions increase susceptibility to infections [46]. Studies on cosmonauts during space flight have shown that IgG levels were unchanged, whereas IgA and IgM levels were sometimes increased [47]. A decreased cytotoxicity in cosmonauts after space flight can be proposed, and this includes the defective function of NK cells and the reduced number of circulating effector cells [48]. Physical and psychological stress associated with space flight resulted in decreased virus-specific T-cell immunity and reactivation of EBV [49]; almost certainly immunity changes in space are similar to those occurring during acute stress conditions. Therefore, it is reasonable to consider stress-related immunotherapy approaches in the practice of space medicine mainly because concerns have been raised about the possible risks of post-flight infections. A decrease in the number of T-lymphocytes and impairment of their function is an important effect of weightlessness on the immune system. From a nutritional point of view, we have to consider that deficiency of zinc is associated with similar changes in T-lymphocytes. Suggestions that modifications to the diet may have a beneficial effect on health are not new. It is well known from ground research that a lack of macronutrients or selected micronutrients, like zinc, selenium, and the antioxidant vitamins, can have profound effects on immune function [50-52]. Such a lack of nutrients also leads to deregulation of the balanced host response [53]. In fact, several micronutrients such as vitamin A, beta-carotene, folic acid, vitamin B12, vitamin C, riboflavin, iron, selenium, zinc have immunomodulating actions [54, 55]. Recent work demonstrates that some nutrients such as arginine, glutamine, nucleotides and omega-3 fatty acids may affect immune function [56]. The dietetic intake of these nutrients should be considered in order to recommend appropriate nutritional supplementation aimed at reducing immune changes due to sub-optimal nutrition that may eventually have negative consequences on immune status and susceptibility to a variety of pathogens. Detailed information on the effects of many micronutrients during space flight are mandatory before specific nutritional recommendations can be made, especially with respect to their relationship with immune system function.

Moreover, the latest data (from healthy subjects and patients with inflammatory diseases) show that probiotics can be used as innovative tools to improve the intestine's immunologic barrier and produce a gut-stabilizing effect. Many of the probiotic effects are mediated via immune regulation, in particular by control of the balance of proinflammatory and antiinflammatory cytokines [57]. Recent results demonstrate that dietary consumption of specific probiotics can enhance natural immunity in healthy elderly subjects [58]. This evidence suggests that the consumption of cultures of beneficial live microorganisms that act as probiotics should be evaluated in target specific populations, including astronauts.

## 12. Discussion

Adequate nutrition is critical for maintenance of crew health during and after extended-duration space flight. Provision of a variety of available foods [59] with positive sensory characteristics [60], and adequate time for preparation and consumption of meals enhances food intake by the crew members. The observed changes in food intake, hypothalamic monoamines, and peripheral hormones suggest that besides microgravity, continuous light exposure contributes to the observed anorexia, and its metabolic sequelae including bone loss [61]. Poor or inadequate sleep may affect eating and drinking behavior, thus, generating the potential for nutritional problems.

The impact of weightlessness on human physiology is profound, with effects on many systems related to nutrition, including bone, muscle, hematology, fluid and electrolyte regulation. Additionally, we have much to learn regarding the impact of weightlessness on absorption, metabolism, and excretion of nutrients, and this will ultimately determine the nutrient requirements for extended-duration space flight. Existing nutritional requirements for extended-duration space flight have been formulated based on limited flight research, and extrapolation from ground-based research. In this regard, NASA's Nutritional Biochemistry Laboratory was charged with defining the nutritional requirements for space flight. This is accomplished through both operational and research projects. A nutritional status assessment program is included operationally for all International Space Station astronauts. This medical requirement includes biochemical and dietary assessments, and is completed before, during, and after the missions. This program will provide information about crew health and nutritional status, and will also provide assessments of countermeasure efficacy.

Although experience with long-term space flight has provided considerable confidence in the ability of the human body to recover from space flight and readapt to the Earth environment, effects observed on the long Skylab, Mir, and NASA-Mir missions have convinced flight physicians and scientists that countermeasures and monitoring are essential to the success of long duration spaceflight. Countermeasures are methods used to limit the negative physical and psychological effects of the space environment on humans. Nutrition and foods are essential for maintenance of health and for enabling certain countermeasures such as exercise. Critical questions and a road map point to the important areas of nutrition and food science research needed in the future. These include development and use of genomic and proteomic research and development and use of other advanced technologies. There must be interactions between the various disciplines to determine the underlying mechanisms and to apply them to nutritional requirements to ensure a healthy and productive crew. However, due to limitations of spaceflight research opportunities, ground-based models are essential for understanding nutrition-related physiologic changes and their underlying mechanisms. Nutrition research efforts require a wide range of models and interdisciplinary approaches including contributions from physiology, biochemistry, psychology, food science and technology, horticulture, and advanced medical technologies.

**13. Conclusions**

After 55 years of human spaceflight from the first human Yuri Gagarin to journey into outer space and a great deal of space life sciences research, much has been learned about human adaptation to microgravity exposure.

NASA has an exciting new vision of future spaceflight to favour the return of humans to the moon in preparation for visits to Mars and possibly beyond. Moon missions are essential to the exploration of more distant worlds. Extended lunar stays build the experience and expertise needed for the long-term space missions required to visit other planets.

Moreover NASA is working toward sending astronauts to a near-Earth asteroid by 2025, then on to the vicinity of the Red Planet by the mid-2030s. Part of this preparation involves studying the psychological and physiological effects of long-term spaceflight, which the agency investigated during 2015 on one-year missions to the International Space Station (The standard stay for astronauts aboard the orbiting lab has been six months).

The biggest challenge to human exploration of deep-space destinations is related to high radiation levels beyond Earth orbit. With current spacecraft technology, astronauts can cruise through deep space for a maximum of one year or so before accumulating a dangerously high radiation dose, researchers say. As a result, many intriguing solar system targets remain off-limits to human exploration at the moment. A one-year spaceflight cap would still allow manned missions to some intriguing destinations, such as Mars.

For future long-term mission replacement of gravity by centrifugation ("artificial gravity") has been proposed as a multi-system countermeasure [102, 103], but particularly for bone. Although some initial studies have been completed, the optimal artificial gravity prescription for bone, including dose, duration, and frequency of centrifugation, remains to be identified [82, 104], along with its potential impact on nutrition and related systems [105, 106].

In fact, data gathered by NASA's Curiosity rover suggest that astronauts could endure a six-month outbound flight, a 600-day stay on the Martian surface and the six-month journey home without accumulating a worryingly high radiation dose.

Mars missions will require additional technological and biomedical advances. Current scenarios, based on existing propulsion technology, are for missions of about 1-1.5 years, with 6 months of transit to Mars, 2-6 month stay on the Mars surface and another 6-month voyage to return. On these flights, early return will not be possible, and thus in situ medical capabilities are needed. Determining what diagnostic testing is required and developing technologies to allow such testing in microgravity or 1/3-gravity will be challenging, to say the least.

It can be realistically predicted that nutritional balance and dietary adequacy will become increasingly important on future long-term space flights especially regarding future missions to the International Space Station and to Mars.

From a nutrition perspective, critical questions remain regarding the nutrient requirements for extended-duration missions and the ability of nutrients to serve as countermeasures to mitigate some of the negative effects of spaceflight. Nutrition is an essential part of maintaining the endocrine and immune system, skeletal and muscle integrity, and the hydration status of the space crew, all of which are important for extended-duration missions and in those where strenuous extravehicular activities are required. In addition, the psychological aspects of individuals during space flights underline the role of the mealtime perceived as a welcome break from job-related activities as well as a chance to socialize.

Initial studies are underway to better understand nutritional requirements in microgravity, the stability of nutrients in foods stored in space, and oxidative damage and how to counteract it. Moreover the design of future space suits is currently underway, and one of the considerations is expanding the ability to provide nutrition during extravehicular activity, either by making a nutritional beverage available in the suit or by making it possible for an astronaut to easily exit the suit for a snack or lunch.

From a food perspective, storage of foods for up to 5 years will be required (as much of the food as possible will be sent ahead of the crewed mission), and ensuring adequate nutrient content at the time of consumption will be critical. For astronauts depending for months to years on a closed food system, any nutrient deficiency or excess could be catastrophic.

Diet will be important for the efficacy of the countermeasures and prevention of compounding problems. Further efforts will be implemented to ameliorate dietary adequacy and food safety in order to counteract deleterious physiological changes, psychosocial repercussions and microbiological hazards. Finally, all space food systems will be designed to meet not only individual nutritional requirements but also different food preferences and eating habits, according to the different cultural background and country of origin of the crew members.

The question of in situ production of food based on crops will be the solution and the additional value to provide a set of nutrients for crew members. Vitamins A, B12, C, D, E, K, etc. as well as mineral salts could be used in future as countermeasures for the radiation exposure of deep space during long-duration space missions. Genetically modified foods could also be the key to solve sudden nutritional problems during long-term missions. Nutrition will be the key-factor for the next phases of exploration beyond this planet, and space nutrition will ensure optimal health and mission success.